\documentclass[12pt]{article} 
\usepackage{fancyvrb,graphicx,amsmath,url,setspace}
\usepackage[font=small,format=plain,labelfont=bf,up,textfont=it,up]{caption}

\begin{document}

\title{\textbf{Unusual ferromagnetism in nanoparticles of doped oxides and manganites} \\}
\author{\textbf{\small Vatsal Dwivedi} \\ {\footnotesize Department of Physics} \\ {\footnotesize University of Illinois at Urbana-Champaign, IL 61801, USA} \\ \\
\textbf{\small A. Taraphder} \\ {\footnotesize Department of Physics and Center for Theoretical Studies} \\ {\footnotesize Indian Institute of Technology, Kharagpur 721302, India} \\ \; \\ \; \\ \; \\}
\date{}

\maketitle

\begin{abstract}
The observation of unusually large ferromagnetism in the nanoparticles of doped oxides and 
enhanced ferromagnetic tendencies in manganite nanoparticles have been in focus recently.  
For the transition metal-doped oxide nanoparticles a phenomenological `charge transfer 
ferromagnetism' model is recently proposed by Coey {\it et al.} From a microscopic 
calculation with charge transfer between the defect band and mixed valent dopants, 
acting as reservoir, we show how the unusually high ferromagnetic response develops. The 
puzzle of nanosize-induced ferromagnetic tendencies in manganites is also addressed within 
the same framework where lattice imperfections and uncompensated charges at the surface 
of the nanoparticle are shown to reorganize the surface electronic structures with 
enhanced double exchange. 
\\ \\
\end{abstract} 

\onehalfspacing
\pagebreak

% SECTION 1 : INTRODUCTION
% ================================================================================

\section{Introduction}
In recent years, there has been a flood of reports on ferromagnetic (FM) tendencies 
associated with nanoparticles and thin films. There are two major class of systems
that show this behaviour, the transition metal-doped (transparent) oxides and the 
colossal magnetoresistive (CMR) manganites. The bulk samples of many of the former  
systems are nonmagnetic, e.g. CuO, TiO$_2$ while the manganites are antiferromagnetic
(AFM) with charge and orbital order.  

One aspect of this FM tendency is that it is almost certainly linked to 
the inhomogeneities, for example, the dopant cations for the oxides and the 
surface states and/or the defects in the nanosize materials for manganites. The FM 
tendency is not present in well-crystallized bulk samples. Secondly, the FM order is 
present in only a fraction of the 
sample volume. In doped oxides, the magnetization, however, is much higher 
than would come from the 
dopants alone. In view of this strange observation, it has been argued that the 
effect is not just an `impurity effect', rather the dopant impurities must in some
way dramatically modify the electronic organization of the entire system (or a 
considerable part thereof). If a fraction of the lattice sites are used to explain 
the magnetic effects, it is not 
necessary that the fraction of electrons in those sites only are responsible  
for the observed magnetic properties: there can be transfer of electrons from 
other sites to the sites in question, namely, the surface, interfaces or the defects. 

Such a phenomenological idea has been proposed~\cite{coey1,coey2} by Coey {\it et al.} to 
account for the high temperature FM in transition metal doped oxides (e.g., Fe:TiO$_2$, 
Fe:CuO and others). The key idea here is that the magnetism is coming from the regions of 
defects (like phase segregated regions, stripes, twin boundaries) coupled to 
the dopants. Electrons in the defect band are  
coupled with the mixed valent dopant cations transferring electrons between 
the two subsystems. The chemical potential of the correlated defect band 
is tuned by the dopant electrons (which form a `\emph{reservoir}' level). It is possible that 
depending on the position of the chemical potential with respect to the van Hove 
singularity of the defect band, a Stoner type instability can be tuned leading to 
a high temperature magnetic long range order (LRO). In particular, if the chemical 
potential is close to the band singularity, the Stoner criterion $U\rho(\varepsilon_F)=1$ 
is easily satisfied, where $U$ is the local Coulomb repulsion in the defect band.  

In order to make these ideas more concrete, we take a locally correlated itinerant 
electron system coupled to a reservoir, which can transfer electrons or holes to 
the band. For the doped oxides, the itinerant band is formed by delocalization of 
electrons in the defect bands~\cite{coey1,coey2} as discussed above. In the 
manganite nanoparticles, the surface electrons form the analogue of the `defect' band 
while the imperfections, broken/dangling bonds and excess charges present at  
the surface~\cite{dagotto,gehring,at} act as the reservoir. The itinerant system 
is derived from the bulk, hence it can be expected to have characteristics of the bulk. 
Its chemical potential is dictated by the filling. The charge reservoir, owing to 
its ability to transfer electrons (holes), can alter the filling of the band. 
For instance, for low fillings, the reservoir can transfer electrons leading to an  
increase in filling and for higher filling the opposite could happen, depending on  
the location of the reservoir. This could pin the chemical potential of the itinerant  
band close to the peak in the density of states (DOS) leading to an enhancement 
of ferromagnetism. 

We work out such models for two cases. Firstly, the phenomenological calculations 
of Coey {\it et al.} on \emph{charge transfer ferromagnetism} is studied more carefully 
starting from a Hamiltonian and the results compared with their phenomenological 
arguments. This puts the model on a microscopic basis. Second, for the manganites, 
the usual Zener double exchange model is treated in the presence of a reservoir as 
argued above, using Monte Carlo simulations in tandem with exact diagonalization 
of the Fermionic part~\cite{dagotto}. The rest of this paper is organized as follows: 
section 2 describes the microscopic model for charge transfer ferromagnetism, 
and section 3 describes the model and results for manganites. 

\section{Doped oxide nanoparticles}

\subsection{Introduction}
The concept of a charge reservoir in the context of inducing ferromagnetism was 
introduced by Coey {\it et al.} (\emph{charge transfer ferromagnetism} model~\cite{coey1, 
coey2}) to explain the ferromagnetism observed in nanostructures (thin films, 
nanoparticles) of insulating oxides like rutile ($TiO_{2}$)~\cite{coey1} or 
$CuO$~\cite{coey2} doped with 1-5\% of iron. In this case, the charge reservoir is due 
to the multiple oxidation states of iron, which can donate (or accept) an electron via 
the ionization process $ Fe^{2+} \longleftrightarrow Fe^{3+} + e^-$. This reservoir can transfer electrons to the band  
formed by defects, twin boundaries, stripes, phase segregated regions, referred to, in 
general, as the \emph{`defect band'}. Experimentally, M\"{o}ssbauer spectroscopy shows iron in both $Fe^{2+}$ and $Fe^{3+}$ 
oxidation states in these oxides. 

For the defect band, at zero temperature, the condition of ferromagnetism is given by 
the Stoner criterion $ U\rho(\varepsilon_F) = 1 $, where $U$ is the \emph{Stoner 
integral}. Hence, the origin of magnetization proposed by Coey {\it et al.} is the transfer 
of electrons from reservoir to the conduction band, leading to Fermi level moving 
closer to the band singularity and Stoner splitting of the two spin channels of the 
defect band. Using this and a model DOS, they show that charge 
transfer to or from a reservoir into a narrow, defect-related band can give rise 
to the inhomogeneous Stoner-type wandering axis ferromagnetism that qualitatively 
reproduces the unusual magnetic 
properties of these systems —- high Curie temperature, anhysteretic 
temperature-independent magnetization curves, a metallic or insulating 
FM ground state and a moment that may exceed that of the dopant cations. 

\subsection{Model and calculation}
Based on the ideas discussed above, the model can be represented by a Hamiltonian 
of itinerant electrons on a square lattice with on-site Hubbard interaction, 
coupled to a narrow reservoir (width of which is taken zero here). Such a 
Hamiltonian can be written as:
\begin{eqnarray}
\mathcal{H} = \sum_{<i,j>\sigma}{(-t-\delta_{ij}\mu)c_{i\sigma}^{\dagger}c_{j\sigma}} + 
\varepsilon_D\sum_{i\sigma}{d_{i\sigma}^{\dagger}d_{i\sigma}} \nonumber \\
+ V\sum_{i\sigma}{(c_{i\sigma}^{\dagger}d_{i\sigma}+d_{i\sigma}^{\dagger}c_{i\sigma})} + 
U\sum_{i}{n_{i\uparrow}n_{i\downarrow}}\, , 
\end{eqnarray}
\noindent where $c_{i\sigma}/c_{i\sigma}^{\dagger}$ are the annihilation/creation 
operators for the itinerant electrons, $d_{i\sigma}/d_{i\sigma}^{\dagger}$ are the 
same for the reservoir, $U$ is the on-site repulsion (acting only in the defect 
band), $V$ is the coupling to the 
reservoir and $\varepsilon_D$ is the position of the reservoir with respect to the peak   
of the defect band DOS. The parameters of the Hamiltonian are varied to obtain the 
phase diagrams.  Typical values, for example, are given in several reviews~\cite{coey1}.
In order to proceed, we use the mean-field approximation for the Hubbard term 

$$n_{i\uparrow}n_{i\downarrow} = \left<n_{\uparrow}\right>n_{i\downarrow} + 
\left<n_{\downarrow}\right>n_{i\uparrow} - \left<n_{\uparrow}\right>\left<n_{\downarrow}
\right>$$ 

\noindent The self-consistent solutions can be found out easily from  
\begin{align}
\mathcal{H} = & \sum_{<k>\sigma}{\left(\varepsilon_k c_{k\sigma}^{\dagger}c_{k\sigma} + 
\varepsilon_D d_{k\sigma}^{\dagger}d_{k\sigma} + V(c_{k\sigma}^{\dagger}d_{k\sigma} + 
d_{k\sigma}^{\dagger}c_{k\sigma})\right)} \nonumber \\
& \; \; \; + U\sum_{k}{\left(\left<n_{\uparrow}\right> n_{k\downarrow} + \left<n_{\downarrow}\right>n_{k\uparrow}\right)} 
\end{align}
\noindent where, for a square lattice, we employ a tight binding dispersion 
$\varepsilon_k = -2t(\cos k_x + \cos k_y) - \tilde{\mu} \, , \, 
\tilde{\mu} = \mu + \frac{U}{2}$ whose DOS has a weak logarithmic 
singularity at zero energy. The self-consistency equations for $\left<n_{\sigma}\right>$ 
were solved over a momentum grid in the first Brillouin zone till convergence 
within 0.1\% is reached.   
 
\subsection{Results}
\begin{figure}
\centering
\includegraphics[height=3.2in,width=3.5in]{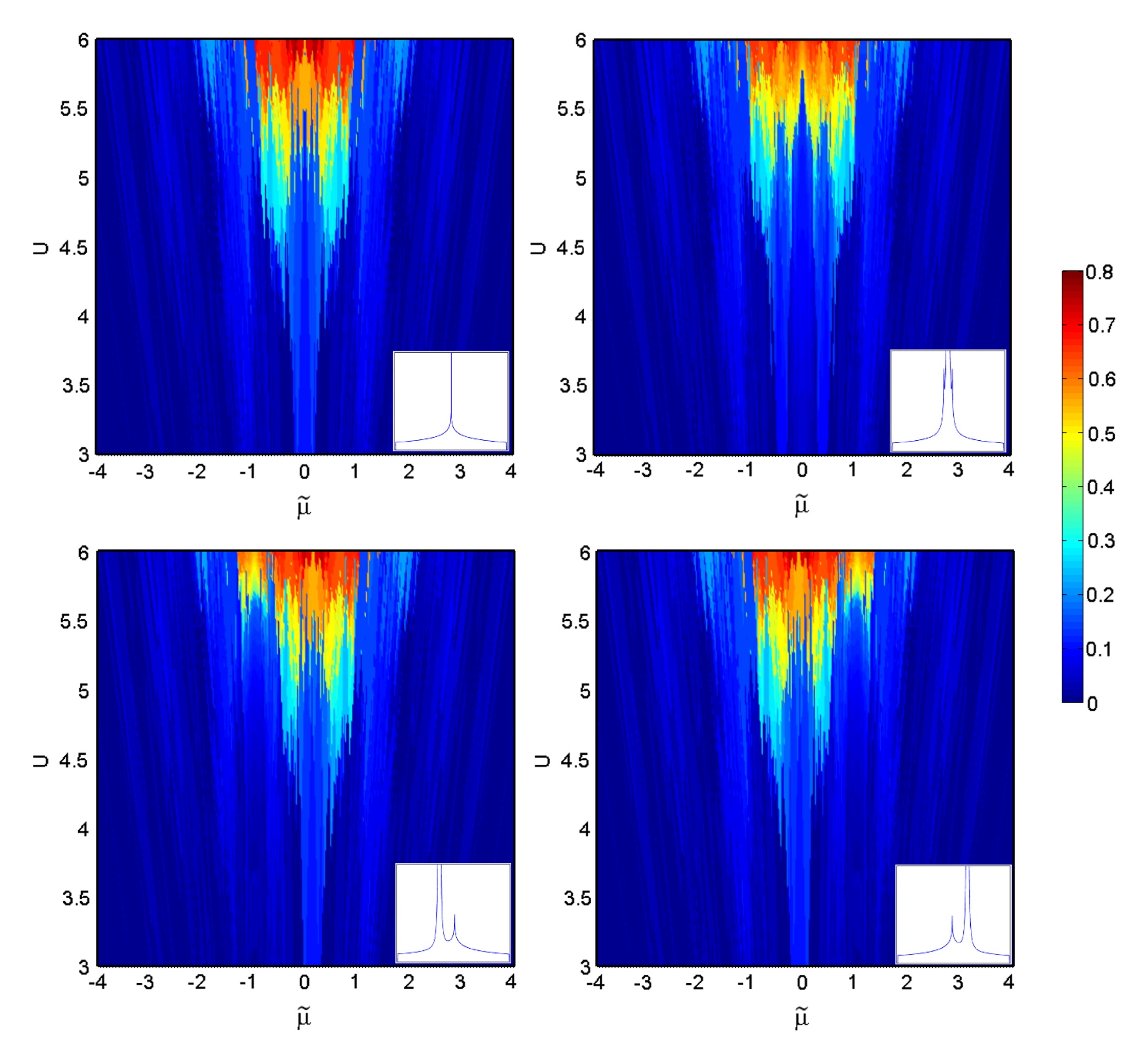}
\caption{The variation of magnetization on the $U-\tilde{\mu}$ plane with (left to right,  
top and bottom column) no reservoir, reservoir at the peak of DOS, reservoir to the left 
and right of the band peak. The color code represents magnetization. Insets in 
each of the figures show the position of the reservoir (broadened for a nonzero $V$) 
with respect to the defect band. There is an enhancement 
of FM (red/yellow regions) when the Fermi level is close to the peak of the 
defect band (shown in insets).} \label{CTFres}
\end{figure}

The magnetization $M = \left<n_{\uparrow}\right> - \left<n_{\downarrow}\right>$ was 
calculated as the Fermi level is moved and for various fixed positions of the reservoir
with respect to the defect band. The result, shown in Fig. \ref{CTFres}, clearly shows 
the effect of adding a reservoir to the system - the magnetization 
is enhanced when the Fermi level is close to the reservoir. It shows how 
ferromagnetic regions in the phase space arise. The location of the FM region in the 
phase space strongly depends on the relative positions of the Fermi energy $\mu$ 
and $\varepsilon_D$. As $\mu$ passes through the reservoir, there is a strong mixing 
of the band and reservoir electrons. The Fermi level gets pinned at the resonant 
level and Stoner criterion is easily met. There is, of course, always a strong 
FM enhancement when the Fermi level is close to the peak of the defect band.  
In the absence of reservoir, the system exhibits a electron-hole symmetry at half-filling, 
as the band is symmetric about the peak. The defect band is expected to be spin-split when the 
Fermi level is close to the peak of the DOS (Stoner splitting), where the characteristics 
of this splitting is symmetric about the peak of the DOS due to the symmetry of the band. 
As discussed above, the effect of reservoir is most pronounced when the reservoir is placed 
off the peak of the DOS, thereby breaking the symmetry in the splitting. A higher splitting 
is seen when the Fermi level is close to the reservoir. 

Note that this mechanism is 
independent of the nature of the underlying lattice. In fact, the special nesting in a 
square lattice generally favors AFM ground state at half-filling. But the mechanism 
discussed here can appear at any filling and the special topology of the 
Fermi surface 
is easily destroyed in a real system by beyond nearest-neighbor hopping. 
 
The high magnetization obtained when the Fermi level is close to the reservoir 
is qualitatively similar to that obtained by Coey {\it et al.} from phenomenological 
calculations using a Lorentzian defect band~\cite{coey1} 
and a Lorentzian band for the reservoir~\cite{unpub}.

% SECTION 3 : MANGANITES
% ================================================================================

\section{Manganites}

\subsection{Introduction}

Manganites came back to focus about seventeen years back owing to the 
discovery~\cite{jin} of colossal magnetoresistance in these compounds. 
These are a 
class of manganese compounds of composition A$_x$B$_{1-x}$MnO$_3$ 
(A,B = La, Ca, Ba, Sr, Pb, Nd, Pr), which crystallize in the cubic structure of 
the perovskite mineral CaTiO$_3$~\cite{tokura}. The basic unit of all the manganites is the 
MnO$_6$ octahedron with corner-shared oxygen and the central Mn$^{3+/4+}$ ion. 

For manganites, the active electronic levels are the 5-fold degenerate \emph{d}-levels 
of the Mn$^{3+/4+}$. In the octahedral environment of MnO$_6$ the $d^5$ is split 
into three-fold degenerate $t_{2g}$ lower level and two-fold degenerate $e_g$ 
upper level. The $t_{2g}$ levels are electronically inert and can be treated as  
localized spins with magnitude $S = 3/2$. These localized spins are coupled to 
the itinerant $e_g$ electrons via Hund's coupling. The itinerant electron system 
forms a band, the filling of which is controlled by the divalent cation doping.

The understanding of the magnetic effects in manganites is governed by the 
\emph{double exchange model} by Zener~\cite{zener}, which 
gives a mechanism for hopping in the $e_g$ levels. The hopping is explained by the 
degeneracy of the Mn$^{3+} \!\!-\! $O$^{2-} \!\!-\! $Mn$^{4+}$ 
and Mn$^{4+} \!\!-\! $O$^{2-} \!\!-\! $Mn$^{3+}$ configurations. It involves a 
simultaneous transfer of an electron from Mn$^{3+}$ to O$^{2-}$ and 
from O$^{2-}$ to Mn$^{4+}$. But as the electrons in the itinerant band are 
coupled to the localized, $t_{2g}$ electrons via a Hund's coupling, the localized 
electron at a site will favour an $e_g$ electron of parallel spin on the site. 

It is well known that manganites have a large Hund's exchange. In the limit 
this is infinite, each of the 5 d-orbitals will be spin split, the `wrong spin' 
orbitals are never populated (as there are only 3 or 4 electrons in 3d
orbitals of Mn ion). In this limit, the double occupancy at each orbital is 
also irrelevant. It suffices to work 
with three degenerate $t_{2g}$ orbitals and one $e_g$ orbital.  

The nanostructured manganites show unusual magnetic behavior, different from the 
bulk. It has been observed in several manganites that the charge ordered, AFM  
manganites, when reduced to nanosize, develop ferromagnetic tendencies 
\cite{rao}, presumably with the charge order also destabilized. Two 
possible scenarios have been put forward to `explain' this: (i) the nanosize 
effectively increases the surface pressure, $P \sim S/R$, where $S$ is 
the surface tension and $R$ is the radius of the nanograin, assumed 
spherical. This excess pressure is supposed to destroy the charge 
order~\cite{tapati}. Pressure induced melting of charge and AF order has 
indeed been seen in bulk manganites. (ii) The enhancement of FM state comes 
from intrinsic causes~\cite{dagotto,gehring,at}. The reconstruction
at the surface of a nanograin reorganizes electronic states and favours double 
exchange. This would favour the FM tendencies over the superexchange between 
Mn ions and, at sufficiently small sizes, completely destabilize AFM order. 
The second view
is emboldened by the observation~\cite{at} that the excess pressure on a typical 
nanograin is about 2-3 GPa, too low to melt the charge and AFM order. 
Besides, recent neutron scattering experiments~\cite{jirak} show no observable 
strain effects in bulk LaCaMnO$_3$ up to about 30 GPa pressure.

\subsection{Model and calculation}
\begin{figure}
\begin{minipage}[b]{0.3\linewidth}
\centering
\includegraphics[height=1.5in,width=2in]{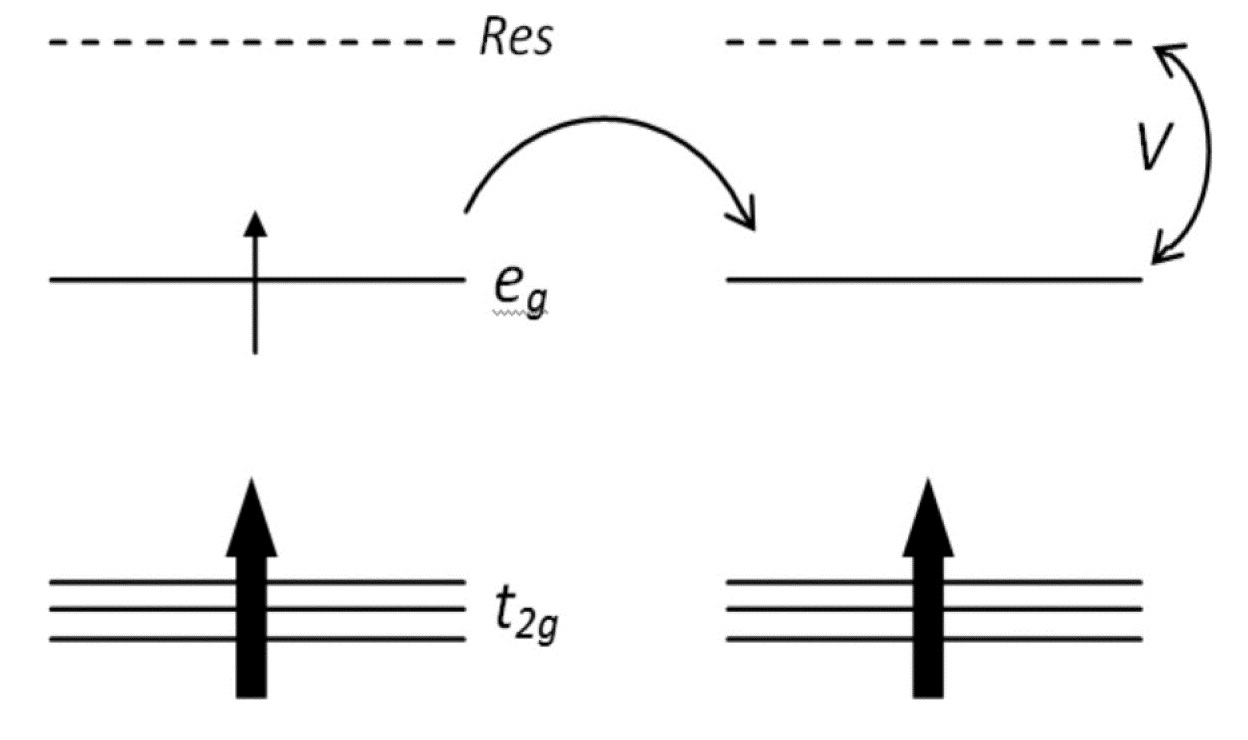}
\caption*{(a)}  
\end{minipage}
\begin{minipage}[b]{0.75\linewidth}
\centering
\includegraphics[height=1.6in,width=4.5in]{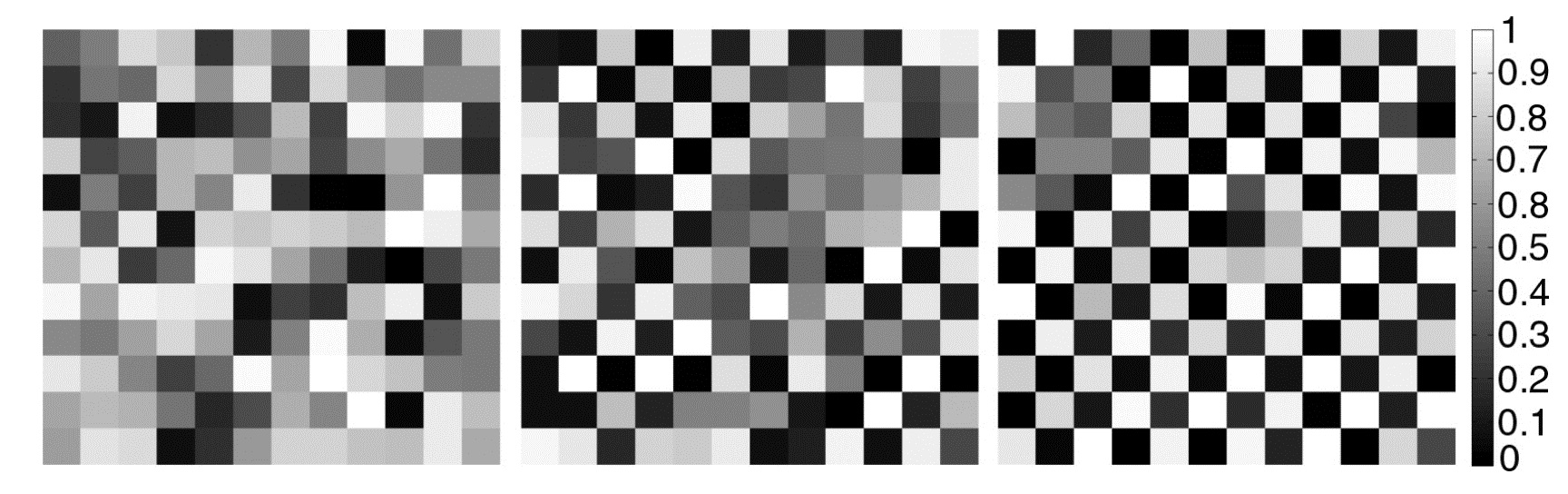}
\caption*{(b)}  
\end{minipage}
\caption{(a) The various levels at each site : The bottom $t_{2g}$ level with a localised classical spin, the middle itinerant $e_g$ level coupled to $t_{2g}$ level via Hund's coupling and the top reservoir level coupled to the middle level. (b)The various states in Monte Carlo simulations, with the localised spin azimuthal angle $\theta_i$ colour-coded in $[0,\pi]$. The figures indicate, from left to right, a ferromagnetic, phase segregated and an antiferromagnetic state, obtained as $J$ is increased for a given $\mu$.}   \label{schm}
\end{figure}

Following the discussions above, we use a single band (coming from the lone $e_g$ orbital) 
Hamiltonian coupled to a reservoir 
at each site. The basic manganite Hamiltonian is an itinerant $e_g$ electron system 
coupled to a localized $t_{2g}$ electrons via Hund's coupling and a superexchange 
interaction between the localized electrons (leading to antiferromagnetism). The localized 
$t_{2g}$ electrons are treated as classical spins of magnitude $S=3/2$ pointing at an 
angle $\theta$ with the spin quantization axis (taken z-axis here). A schematic of 
this model is depicted in Fig. \ref{schm}. The Hund's coupling is taken 
as $J_H \rightarrow \infty$. In this limit, the hopping integral is modified by the 
projection of spin at site $i$ onto its nearest neighbour $j$~\cite{anderson}. 
The overall Hamiltonian is
\begin{eqnarray}
\mathcal{H} = \sum_{<i,j>}{\left(-t\cos{\left(\frac{\theta_i-\theta_j}{2}\right)}-
\delta_{ij}\mu\right)c_i^{\dagger}c_j} + \nonumber\\
V\sum_{i}{(c_i^{\dagger}d_i+d_i^{\dagger}c_i)}   +  \varepsilon_D\sum_{i}{d_i^{\dagger}d_i} + \nonumber\\
\tilde{J}\sum_{<i,j>}{\cos{(\theta_i - \theta_j)}}
\end{eqnarray}
\noindent where $c_{i\sigma}/c_{i\sigma}^{\dagger}$ are the annihilation/creation 
operators for electrons in the band,  $d_{i\sigma}/d_{i\sigma}^{\dagger}$ are the 
operators for the reservoir, $V$ is the coupling to the reservoir, $\varepsilon_D$ 
is the position of the reservoir with respect to the Fermi level and $\tilde{J} = \frac{9}{4}J$ 
is the superexchange parameter. In computation, all parameters are normalized by the hopping parameter $t$.

For this off-diagonal disordered Hamiltonian we use a hybrid Monte Carlo simulation~\cite{dagotto} 
where the Fermionic part was solved by exact diagonalization and the annealing over classical
variables were performed by Metropolis algorithm. The simulations were carried out on 
a $12\times12$ square 
grid with periodic boundary condition. A vector $\Theta = [\theta_1, \theta_2, \dots 
\theta_{N}]$ uniformly distributed in $[0,\pi]$ is chosen to start with, where 
$\theta_i$ is the azimuthal angle of the localized \emph{classical} spin at site $i$. 
At each step, two $\theta_i$'s were modified by a random amount in $\left[-\frac{\pi}{16}, 
\frac{\pi}{16}\right]$ and the Hamiltonian was diagonalized for this new $\Theta$ 
vector. The choice between the new and the old states was done using the 
Metropolis-Hastings algorithm. The system was annealed in this fashion from $\beta = 
1$ to $\beta = 25$ in 100,000 iterations. The spin-spin correlation and free energy 
was averaged out over a further 50,000 iterations.

\subsection{Results}
\begin{figure}
\begin{minipage}[b]{0.5\linewidth}
\centering
\includegraphics[height=2.6in,width=3.5in]{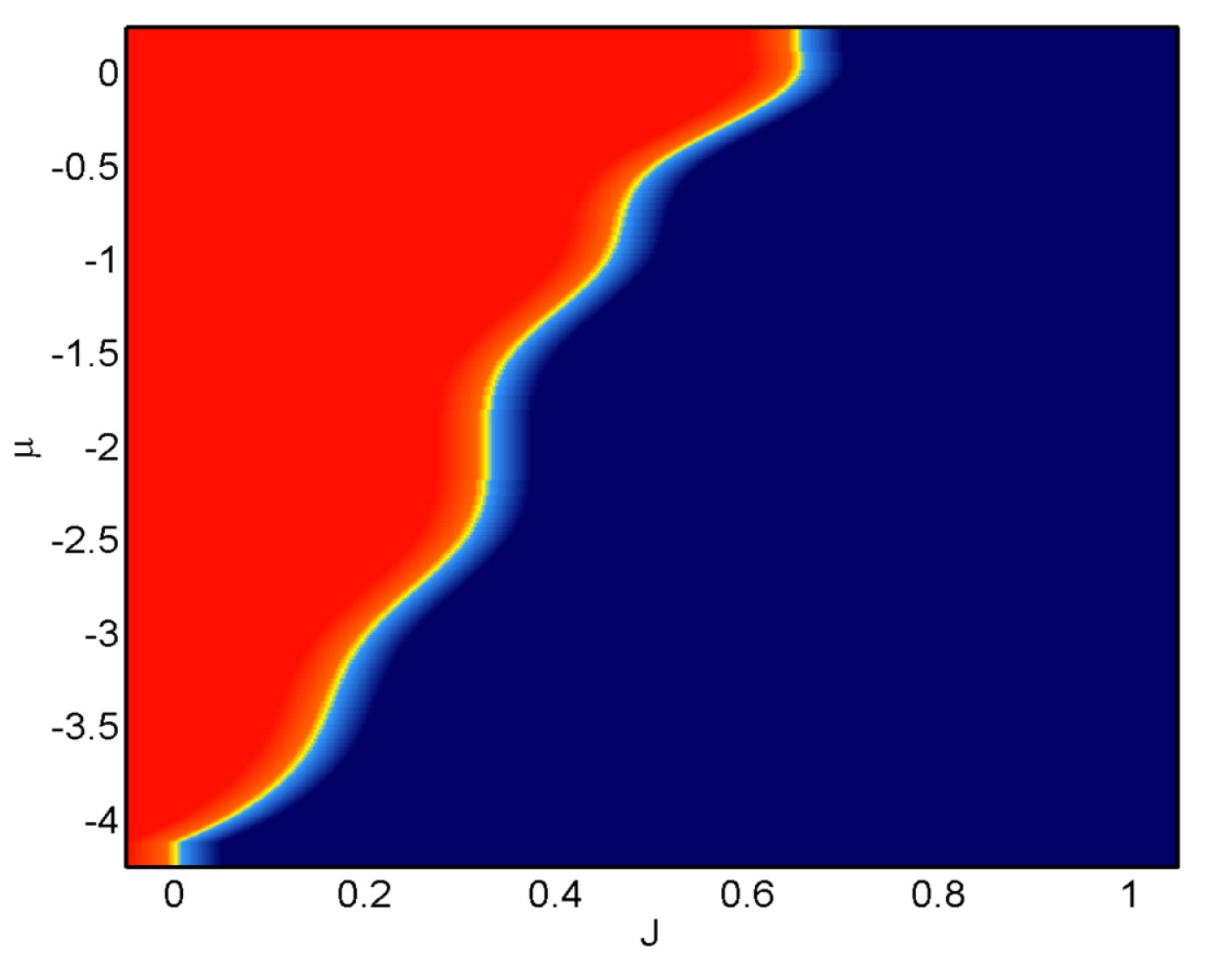}
\caption*{(a)}  
\end{minipage}
\begin{minipage}[b]{0.5\linewidth}
\centering
\includegraphics[height=2.6in,width=3.5in]{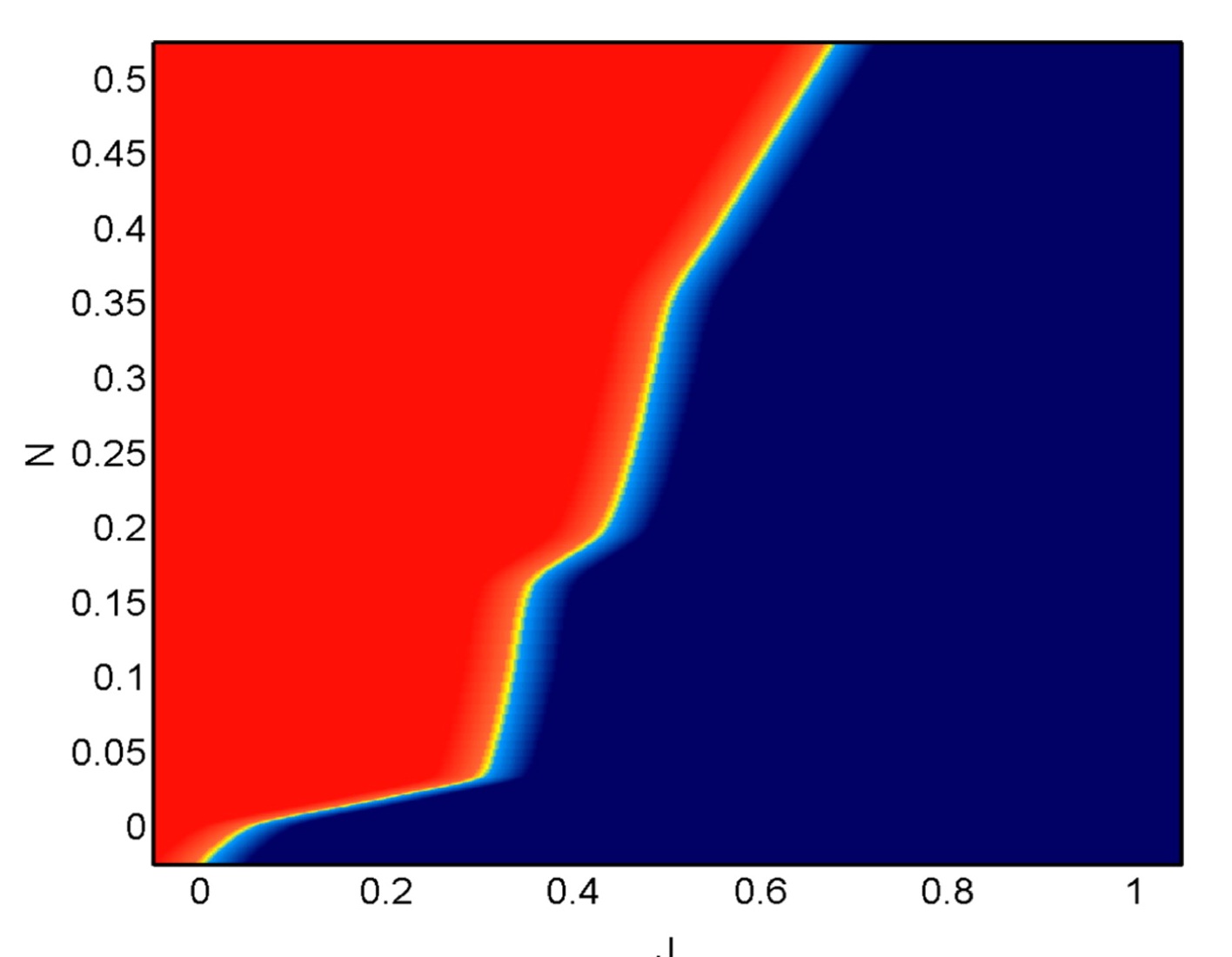}
\caption*{(b)}  
\end{minipage}
\caption{Phase diagram of the manganite system in absence of reservoir on (a)$J-\mu$ and (b)$J-N$ planes. The colour codes are same as in previous figure.}   \label{phase}
\end{figure}

We obtain the results for both with and without the reservoir. In order to obtain the 
phase diagram, simulations were carried out for various parameter values. Typical 
magnetic configurations that appear in the ground states of the MC simulation are 
shown in Fig. \ref{schm}(b). Ferromagnetic, phase segregated and antiferromagnetic 
states are shown. It is not always possible to delineate different phases (particularly 
close to a phase transition). In the theormodynamic limit, there will be a small splitting
between the spin up and spin down bands. However, for the small number of sites we work 
in ($12\times12=144$), a better quantitative method is to find the nearest neighbor spin 
spin correlations, which is -1 and +1 for saturated AFM and FM, respectively. The 
phase boundary is given by $\left<S_iS_j\right> = 0$. The resulting phase diagram is 
plotted in Fig. \ref{phase}(a) and Fig. \ref{phase}(b). The value of $\mu$ is varied 
from $-4$ to $0$, corresponding to zero to half filling, i.e., the value of $x$ varying 
from $0$ to $1$ in, say, La$_{1-x}$Ca$_x$MnO$_3$.

Clearly the presence of a huge number of states leads to a situation where there could
well be degenerate (or nearly degenerate) solutions for the ground state. The competing
interactions of superexchange and double exchange favouring the AFM and FM correlations 
respectively, lead to first order transitions and consequent phase 
segregation~\cite{dagotto}. Indeed, similar situation obtains here too.  
As observed in the simulations, the phase transition occurs via a phase segregated 
state. It is worthwhile to mention the possibility of a \emph{canted spin state} 
~\cite{degen}. From a mean field calculation, it was shown that 
the AFM-FM transition  in the double exchange Hamiltonian should 
proceed through two continuous transitions via a spin canted state. However, 
the canted state has never been found to be the ground state in any 
simulation~\cite{dagotto}. The transition is first order with  
a phase-segregated (two-phase coexistence) region as we also confirm.  

\begin{figure}
\begin{minipage}[b]{0.5\linewidth}
\centering
\includegraphics[height=1.9in,width=3.3in]{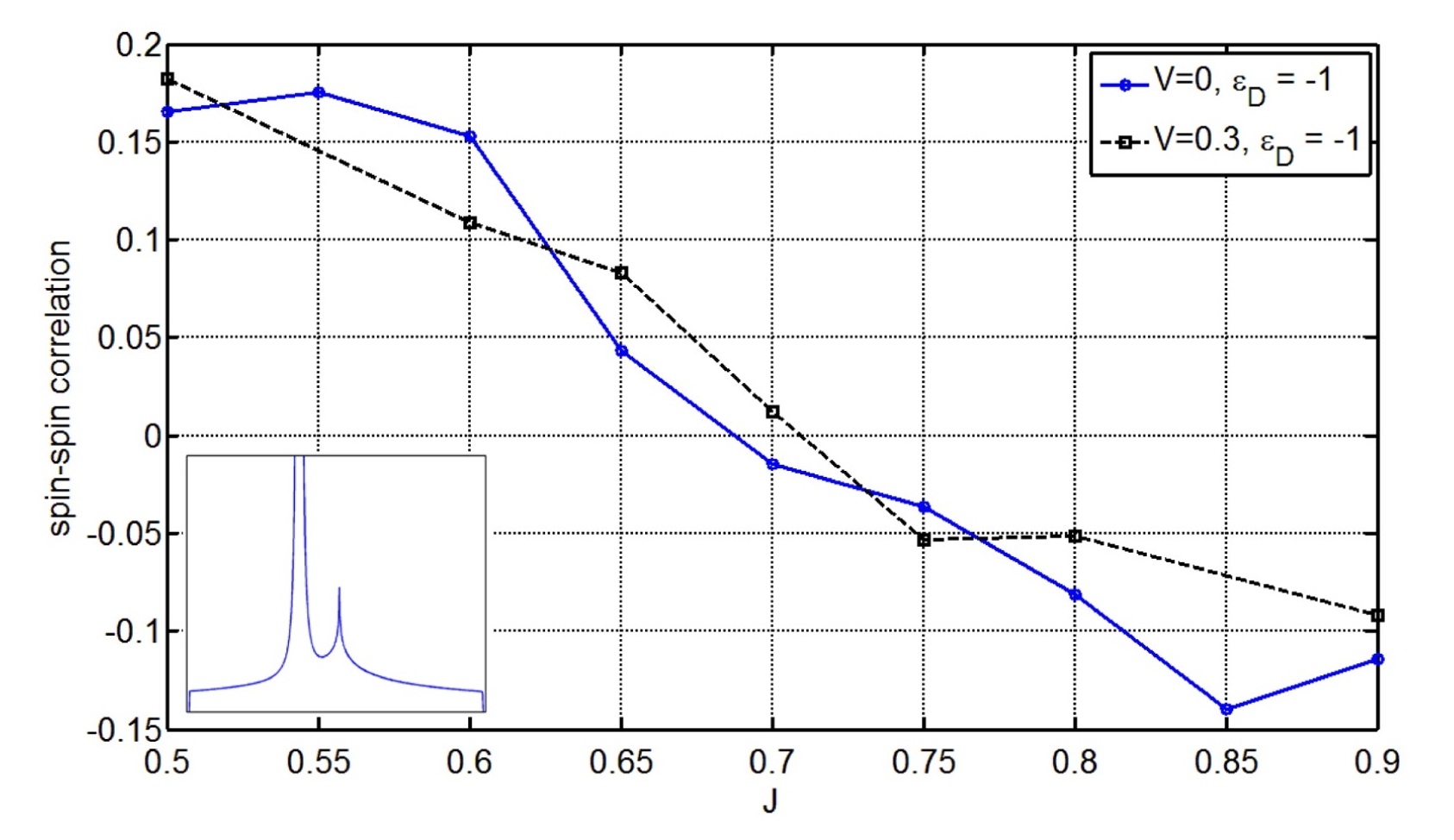}
\caption*{(a)}  
\end{minipage}
\begin{minipage}[b]{0.5\linewidth}
\centering
\includegraphics[height=1.9in,width=3.3in]{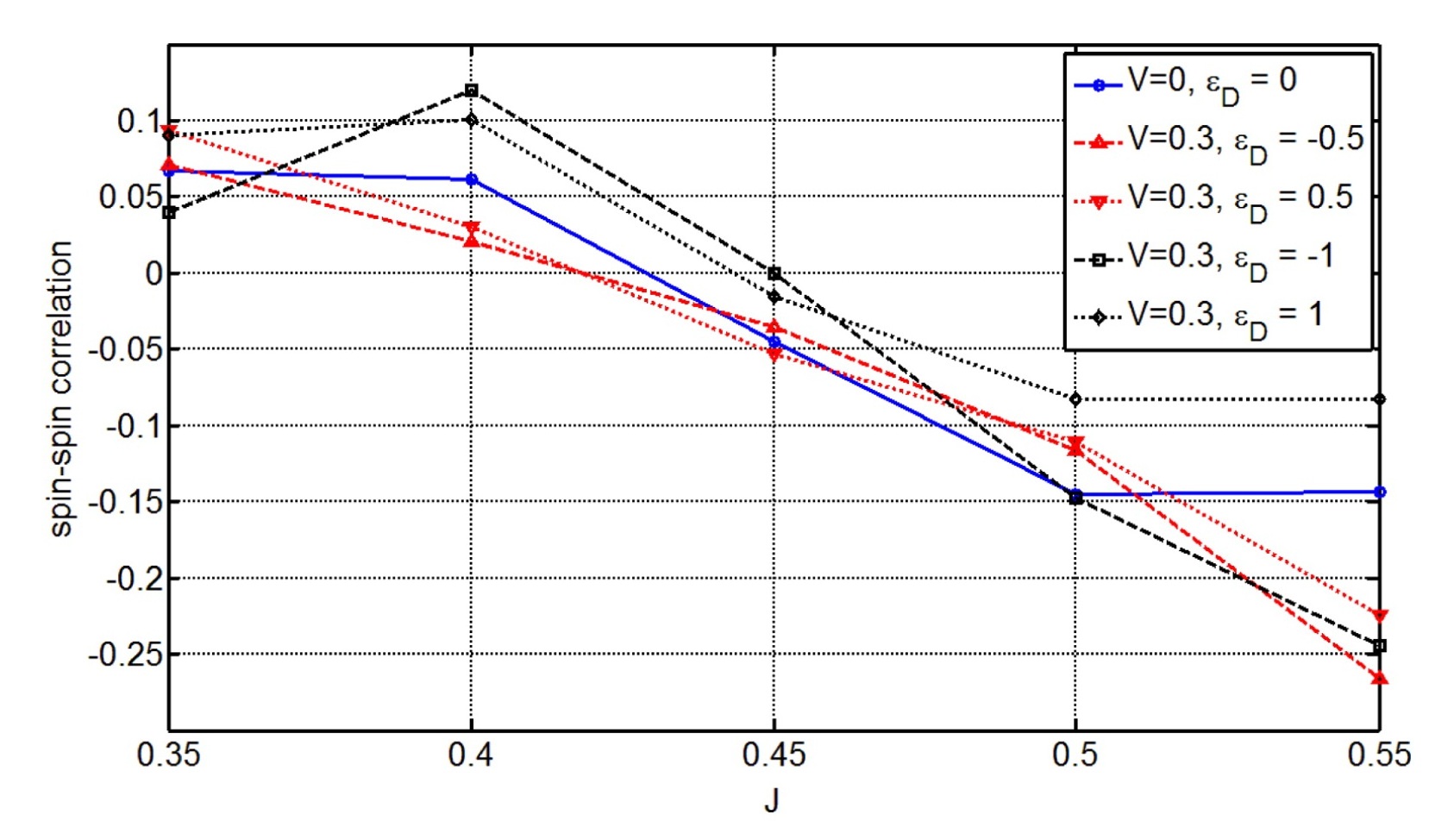}
\caption*{(b)} 
\end{minipage}
\caption{Variation in the spin-spin correlation due to reservoir at (a)$\mu=0$ and (b)$\mu=-1$}  \label{res}
\end{figure}

The addition of reservoir introduces two new parameters, the coupling $V$ and the 
position of the reservoir, $\varepsilon_D$ with respect to the Fermi level. In 
all the calculations, the Fermi level and the position of the reservoir 
are held constant, and 
the coupling is turned on from $V=0$ to $V=0.3$. The spin-spin correlation 
is computed, for the two cases of zero coupling (which is equivalent to no 
reservoir) and with a finite coupling. These calculations 
were repeated for various positions of reservoir with respect to the Fermi level. 
The variation in spin-spin correlation is calculated for $\mu = 0$ (filling = $0.5$) 
and $\mu = -1$ (filling = $0.2$), as plotted in Fig. \ref{res}(a) and \ref{res}(b), 
respectively. 

%\begin{figure}[t]
%\centering
%\includegraphics[height=1.8in,width=3.5in]{res_mu0p.png}
%\caption{Variation in the spin-spin correlation due to reservoir at $\mu=0$. 
%The inset shows the broadened reservoir level when $V = 0.3$.} \label{res0}
%\centering
%\includegraphics[height=1.8in,width=3.5in]{res_mu1p.png}
%\caption{Variation in the spin-spin correlation due to reservoir at $\mu=-1$.} \label{res1}
%\end{figure}

On addition of the reservoir (with $\varepsilon_D=-1.0,\, V=0.3$) to the system below 
the Fermi level at half-filling, the filling of the band decreased from 0.5 to 0.47; 
electrons have been drained 
to the reservoir. At half filling, each site has one electron and delocalization 
is not possible. However, when holes are introduced in this system, the holes 
will delocalize all over the lattice, leading to significant kinetic energy 
gain (enhanced double exchange) and ferromagnetism is enhanced in the ground 
state. As seen in Fig. \ref{res}(a), a comparison between the dashed ($V=0.3$) and 
contunous lines ($V=0$) indicate enhanced FM correlation in the former, indeed higher
spin spin correlation imply more ferromagnetic tendency. The 
corresponding location and broadening of the reservoir is also 
shown in the inset. The presence of the reservoir moves the peak 
of the band towards higher energy by $V^2/t$, thereby leading to decrease 
of the filling of band for $\mu = 0$.     
Hence, the reservoir induces ferromagnetism by introducing holes in the band.
For the second case of $\mu = -1, \,  \varepsilon_D = \pm 1$ (Fig. \ref{res}(b)), 
the reservoir acts as a donor and drives the system towards ferromagnetism again.  
Here, as the original band-filling is low ($0.2$ in this case), 
the delocalization 
energy can be increased only by addition of electrons. The reservoir now acts as 
a donor and accomplishes the same objective as above.

% SECTION 4 : CONCLUSIONS
% ================================================================================

\section{Conclusions}
It is shown that the charge transfer ferromagnetism model introduced by Coey 
et al. is able to predict the hugely enhanced ferromagetic order seen in 
the transition metal doped oxide nanoparticles qualtitatively. More importantly, 
the observed enhancement of FM tendencies in nanoparticles of manganites 
also finds a resolution within the same framework. Surface magnetic probes 
like $\mu$SR may be a possible tool that can establish the proposed scenario 
in manganites.  

% SECTION : REFERENCES
% ================================================================================


\thebibliography{99}
\footnotesize
\bibitem{coey1} J. M. D. Coey, P. Stamenov, R. D. Gunning, M. Venkatesan and K. Paul, New J. Phys. 12 (2010) 053025.
\bibitem{coey2} J. M. D. Coey, Kwanruthai Wongsaprom, J. Alaria and M. Venkatesan, J. Phys. D: Appl. Phys. 41 (2008) 134012.
\bibitem{dagotto} E. Dagotto, T. Hotta and A. Moreo, Phys. Rep. 344 (2001) 1.
\bibitem{gehring} H. Zenia, G. A. Gehring, G. Banach, and W. M. Temmerman, Phys. Rev. B 71 (2005) 024416.
\bibitem{at} S. Kundu, T. K. Nath, A. K. Nigam, T. Maitra and A. Taraphder, 
arXiv:1006.2943 (2010): Jour. Nanoscience and Nanotech. (2011) {\em to be published}.
\bibitem{unpub} Vatsal Dwivedi, P. Stamenov, C. Porter and J. M. D. Coey, Unpublished results.
\bibitem{jin} S. Jin, T. H. Tiefel, M. McCormack, R. A. Fastnacht, R. Ramesh and L. H. Chen, Science 264 (1994) 413.
\bibitem{tokura} Colossal Magnetoresistive Oxides edited by Y.Tokura, Gordon and Breach Science, Singapore, 2000.
\bibitem{rao} S. S. Rao, K. N. Anuradha, S. Sarangi, and S. V. Bhat, Appl. Phys. Lett. 87, 182503 (2005).
\bibitem{tapati} Tapati Sarkar, A. K. Raychaudhuri, and Tapan Chatterji, Appl. Phys. Lett. 92, 123104 (2008).
\bibitem{jirak} Z. Jir\'{a}k, E. Hadov\'{a}, O. Kaman, K. Kn\'{\i}\v{z}ek, M. Mary\v{s}ko, E. Pollert, M. Dlouh\'{a} and S. Vratislav, Phys. Rev. B 81 (2010) 024403.
\bibitem{zener} C. Zener and R. R. Heikes, Rev. Mod. Phys. 25 (1953) 191.
\bibitem{anderson} P. W. Anderson and H. Hasegawa, Phys. Rev. 100 (1955) 675.
\bibitem{degen} P. G. de Gennes, Phys. Rev. 118 (1960) 141.
\end{document}